\documentstyle[PASJadd,psfig]{PASJ95}
\newcommand{\vol}{51}
\newcommand{\no}{6}
\newcommand{\lett}{}
\newcommand{\spage}{}
\newcommand{\rdate}{1999 August 16}
\newcommand{\adate}{1999 September 27}
\begin{document}
\title{ASCA Discovery of a Be X-Ray Pulsar in the SMC: \\  AX J0051$-$733}
\author{Kensuke {\sc Imanishi}, Jun {\sc Yokogawa}, 
Masahiro {\sc Tsujimoto}, 
and Katsuji {\sc Koyama}\thanks{CREST, Japan Science and Technology Corporation
 (JST), 4-1-8 Honmachi, Kawaguchi, Saitama 332-0012.} \\ 
{\it Department of Physics, Graduate School of Science, Kyoto University, 
Sakyo-ku, Kyoto, 606-8502} \\
{\it E-mail(KI): kensuke@cr.scphys.kyoto-u.ac.jp}}
\abst{
ASCA observed the central region of the Small Magellanic Cloud, 
and found a hard X-ray source, AX J0051$-$733, at the position
of the ROSAT source RX J0050.8$-$7316, 
which has an optical counterpart of a Be star. 
Coherent X-ray pulsations of 
323.1 $\pm$ 0.3 s were discovered from AX J0051$-$733. 
The pulse profile shows several sub-peaks in 
the soft (0.7--2.0 keV) X-ray band, but becomes nearly sinusoidal
in the harder (2.0--7.0 keV) X-ray band. 
The X-ray spectrum was found to be hard, and is  
well fitted by a power-law model with a photon index of 
1.0 $\pm$ 0.4.
The long-term flux history was examined with the archival data
of Einstein observatory and ROSAT; 
a flux variability with a factor $\gtsim$ 10 was found.
}
\kword{binaries: general --- pulsars: individual (AX J0051$-$733) --- 
X-rays: general}

\maketitle
\thispagestyle{headings}

%section 1
\section{Introduction}

Systematic X-ray source surveys in the Small Magellanic Cloud (SMC) 
started with the Einstein Imaging Proportional Counter 
(IPC: see e.g.\ Wang, Wu\ 1992), followed by the ROSAT 
Position Sensitive Proportional Counter 
(PSPC: Kahabka, Pietsch\ 1996; Kahabka et al.\ 1999).
Optical identifications, however, require  more accurate determinations 
of the X-ray source positions, and have been made
by Cowley et al.\ (1997) and Schmidtke et al.\ (1999) 
with the ROSAT High-Resolution Imager (HRI). 
They conducted optical photometry and spectroscopy 
studies with the Cerro Tololo Interamerican Observatory (CTIO) 
and identified X-ray sources with optical counterparts. 
Most of these were proven to be supernova remnants (SNRs) in the SMC, or 
foreground bright stars in our Galaxy, 
while four were found to have Be star counterparts within their error regions.
Hughes and Smith (1994) also made an optical counterpart search for
selected X-ray sources, and found two Be star candidates  at the X-ray
positions.

X-ray binary pulsars with a Be star companion (here Be-XBPs) are 
thought to comprise the majority of X-ray binary pulsars (XBPs).  
However, possibly due to their transient nature, 
most of the Be-XBPs have not yet been discovered in the past X-ray
observations. 
The SMC region would not be an exception.  Since a pulsation search from 
X-ray sources with a Be star companion is a direct approach to discover
Be-XBPs, we have performed timing studies for X-ray 
sources with a Be star companion reported 
by Cowley et al.\ (1997),  Schmidtke et al.\ (1999) and Hughes, Smith\ (1994).
The studies were made using archival data of 
the Japanese X-ray satellite ASCA. 

In this Letter, we report on the discovery of X-ray pulsations from 
AX J0051$-$733 = RX J0050.8$-$7316, 
which was the brightest X-ray source among the optically identified 
Be stars.  We also report on the long-term flux variability from Einstein, 
ROSAT and ASCA observations. 

%section 2
\section{Observation}
AX J0051$-$733 was included in the
ASCA observation of the radio supernova remnant N19, on 
1997 November 13--14, pointing at the direction of
(00$^{\rm h}$47$^{\rm m}$16$^{\rm s}$, $-$73$^{\circ}$08$'$30$''$)
 (here and after, we use J2000 coordinates).

ASCA (Tanaka et al.\ 1994) is equipped with four identical X-ray telescopes 
(XRTs: Serlemitsos et al.\ 1995) 
sensitive to photons with energies 0.4--10 keV. 
Four independent detectors,
the two Solid-state Imaging Spectrometers (SISs: Burke et al.\ 1991) 
and the two Gas Imaging Spectrometers 
(GISs: Ohashi et al.\ 1996; Makishima et al.\ 1996), 
are placed at the foci of XRTs and are operated in parallel; 
hence, four independent data sets can be provided. 
However, we analyzed only the GIS data, 
because AX J0051$-$733 was outside of the SIS field of view.

Each of the GIS detectors was operated in the PH mode with the standard
bit-assignment that provides a time resolution of 62.5 ms and 0.5 s in
high and medium bit-rate, respectively. 
Data taken at a geomagnetic cutoff rigidity lower than 4 GV, 
at an elevation angle less than 5$^\circ$ from the Earth and during passage
through the South Atlantic Anomaly were rejected. 
The particle events were also removed by a rise-time discrimination method. 
After these filterings, 
the total available exposure time of each GIS was $\approx 43$ ks.

%section 3
\section{Analysis and Results}

%section 3.1
\subsection{X-Ray Images and Source Identification}

We made GIS2 and 3 images of the N19 region, and found
a bright point source at an off-axis angle of $\sim$ 16$'$ in the GIS 
fields. 
The position was determined to be  (00$^{\rm h}$50$^{\rm m}$50$^{\rm s}$, 
$-$73$^{\circ}$16$'$04$''$), with a possible error of $\sim 1'$.
The larger  position error than the nominal value of 
$\sim 40''$ (Makishima et al.\ 1996) is due to its large off-axis angle. 
Within this error region of AX J0051$-$733, 
we found a ROSAT source, RX J0050.8$-$7316, with coordinates of 
(00$^{\rm h}$50$^{\rm m}$45$.\hspace{-3pt}^{\rm s}$3, 
$-$73$^{\circ}$15$'$54$''$), 
which was optically identified to be a Be star of $V$ $\sim$ 15.4 mag, 
and $B-V$ = $-$0.04 mag (Cowley et al.\ 1997).

%section 3.2
\subsection{Timing Analysis}

Since the shape of the point-spread function is elongated 
at an off-axis position, like AX J0051$-$733, we extracted
photon events for GIS 2 and 3, each from an elliptical region 
with major and minor axes of $\sim$ 5$'$ and $\sim$ 2$'$, respectively.

After a barycentric photon arrival time correction, we performed 
 FFT (Fast Fourier Transformation) analyses; we found a peak in the 
power spectrum.  To maximize the signal-to-noise ratio of this peak, we
tried FFT analyses with many trial energy bands.  Figure 1 shows the
best power spectrum using the 1.5--6.0 keV energy band. 
A total of 541 (GIS 2) and 488 (GIS 3) photons were used for this FFT
analysis.

A clear peak is noted at a frequency of 3.095$\times$10$^{-3}$ Hz, 
corresponding to a pulsation period of $\sim$ 323 s.
The chance probability to detect such a strong 
power signal at any frequency given in figure 1
is only $\sim$ 10$^{-10}$; hence, we can safely 
conclude that the peak is not due to 
Poisson fluctuations, but is a real signature.

To estimate the pulsation period more precisely, 
the epoch-folding technique was applied, 
assuming that no period change occurred during the observation. 
The most likely period was determined to be $P$ = 323.1 $\pm$ 0.3 s.
Figure 2 shows the folded light curves in the soft (0.7--2.0 keV: upper panel) 
and hard (2.0--7.0 keV: lower panel) X-ray bands. 
The pulse profiles are different between these two bands; 
we can see three or four sub-peaks in the soft band, while 
a quasi-sinusoidal curve can bee seen in the hard band. 
The pulse fractions, defined as (Pulsed intensity)/(Total intensity), 
are  $\sim$ 56\% and $\sim$ 80\% in the soft and hard bands, respectively.

\begin{figure}
%\begin{center}
\psfig{file=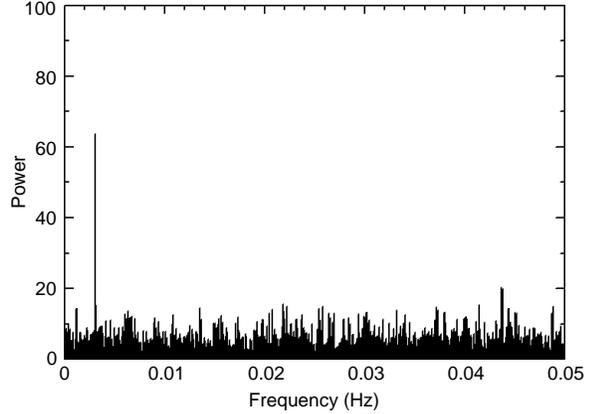,width=0.9\columnwidth}
%\end{center}
\caption{Power spectrum of AX J0051$-$733 in the 1.5--6.0 keV band. 
A significant power can be seen at 0.03095 Hz $\approx$ 323 s.}
\end{figure}

\begin{figure}
%\begin{center}
\psfig{file=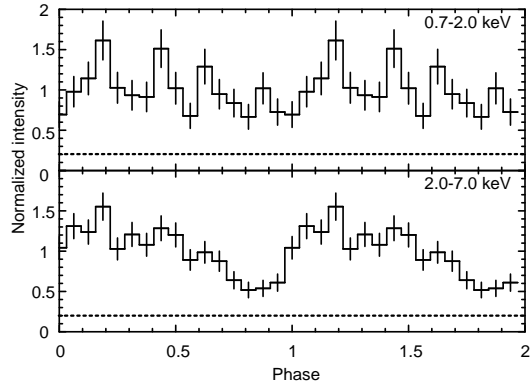,width=0.9\columnwidth}
%\end{center}
\caption
{Folded light curves of AX J0051$-$733 in the 0.7--2.0 keV (upper panel) 
and 2.0--7.0 keV (lower panel) bands.
The background levels are represented by the dotted lines. }
\end{figure}

%section 3.3
\subsection{Spectral Analysis}

Unfortunately, AX J0051$-$733 is located only $\sim 6'$ away from 
the calibration isotope of GIS 3.  We therefore did  not use the GIS 3 
data for the spectral study in order to avoid possible contamination
from the isotope. 
We extracted a source spectrum of GIS 2 from the same region 
as that used for the timing analysis 
and selected a section of nearby sky as a background region. 
Figure 3 shows a phase-averaged source spectrum. 
Since the spectrum showed neither an emission line nor any other prominent
structure, we applied a simple power-law model with interstellar
absorption for the fitting. 
The best-fit model and parameters are given in 
figure 3 and table 1, respectively. 

To determine any possible spectral changes during the pulse, 
we divided the data  into two phases, 
phase 0.0--0.5 (high state) and 0.5--1.0 (low state), 
and fitted two separate spectra. 
Since the statistics of these spectra were limited, we fitted them 
to a power-law model fixing 
the absorption column density to the best-fit value of the phase-averaged 
spectrum.  The best-fit parameters are  summarized in table 1.
The large errors make it unclear whether or not the photon index 
changed with the pulse phase. 

\begin{figure}
%\begin{center}
\psfig{file=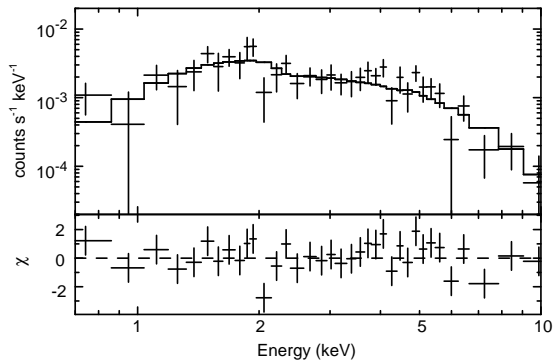,width=0.9\columnwidth}
%\end{center}
\caption
{{\bf Upper panel}: Phase-averaged energy spectrum of AX J0051$-$733 
obtained from GIS2.
The solid line shows the best-fit power-law model.
{\bf Lower panel}: Each residual from the best-fit model.}
\end{figure}

%section 4
\section{Long-Term Flux History}

To investigate any long-term flux variability, we accessed the
HEASARC archive system for the Einstein and ROSAT observations, 
and found that AX J0051$-$733 was included in 1 Einstein IPC, 9 ROSAT HRI 
and 6 ROSAT PSPC observations. 
For each observation, we extracted the source and background events 
from a circular region of 1$'$ radius and an outer region 
of 1$'$--1.\hspace{-3pt}$'$4 radius, respectively. 
After background subtraction, we estimated
the source flux using PIMMS software, 
assuming that the spectral shapes were the same as that obtained with ASCA. 

In figure 4, we summarize the X-ray fluxes from AX J0051$-$733. 
The ASCA flux at MJD = 50765.2--50766.3 was
higher than those of Einstein obtained at MJD = 44189.1--44189.9 and 
those of ROSAT at 49082.2--49104.0  by a factor $\gtsim$ 10.
Therefore, we conclude that AX J0051$-$733 has been highly variable, although 
a possibility that low flux periods are due to eclipses is not ruled
out. 

\begin{figure}
%\begin{center}
\psfig{file=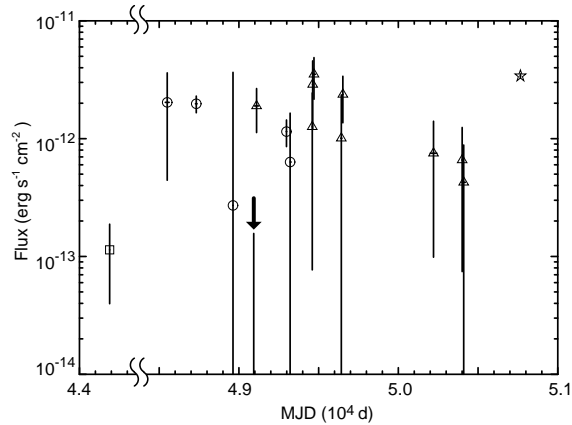,width=0.9\columnwidth}
%\end{center}
\caption
{Flux variation history of AX J0051$-$733. 
Data are obtained by the Einstein IPC (square), ROSAT PSPC (circle), 
ROSAT HRI (triangle), and ASCA GIS (star).
The data point given by the arrow (MJD = 49082.2--49104.0) is the ROSAT
PSPC observation that showed no excess X-ray from AX J0051$-$733, 
hence we can only determine the upper limit of the flux. }
\end{figure}

%section 5
\section{Discussion}
%section 5.1
\subsection{Is AX J0051$-$733 a Be Star Pulsar ?}

We have discovered coherent X-ray pulsations from a Be star binary 
AX J0051$-$733 = RX J0050.8$-$7316.
The long pulsation period ($\sim$ 323 s), large flux variation 
in the long time span, flat power-law spectrum (photon index $\sim$ 1) 
and existence of a Be star companion
strongly support that AX J0051$-$733 is a Be-XBP (Nagase\ 1989).

In the long-term history, we found 7 X-ray outbursts with flux  
$\gtsim$ 2 $\times$ 10$^{-12}$ erg s$^{-1}$ cm$^{-2}$ .  
Most of the known Be-XBPs exhibit highly eccentric orbits, 
and flaring activities are often taken place when a neutron star passes 
near the peri-astron or across the equatorial plane, 
where the circumstellar gas density is expected to be high.
We suspect that the occasional flares from AX J0051$-$733 are related 
to a possible eccentric orbital period.  
However, no clear periodicity is noted in these flare intervals. 
We, nevertheless, folded the long-term light curve with a period
of 185 days, which is near the expected orbital period 
based on Corbet's empirical relation between the orbital and pulse periods 
in a Be-XBP (Corbet\ 1984).  
Although a hint of clustering of the outburst around an interval of 
about 185 d was found, as is given in figure 5, 
no definite argument for an orbital period was obtained. 
To establish the Be-star pulsar binary scenario and its structure, 
an orbital period determination is essential. 
We encourage further monitoring observations of AX J0051$-$733.

%section 5.2
\subsection{Pulsation Search from the Other Be Star Binary Candidates}

The discovery of coherent pulsations from AX J0051$-$733, 
which was optically identified to be a Be star binary in the SMC,
leads us to suspect that the other Be star binary candidates
(RX J0051.9$-$7311, RX J0052.9$-$7158, and RX J0058.2$-$7231: 
see Schmidtke et al.\ 1999; RX J0103.2$-$7209 and RX J0106.2$-$7205: 
see Hughes, Smith\ 1994) also exhibit coherent pulsations (Be-XBPs).
We thus extended our pulsation search to these sources.  

RX J0051.9$-$7311 and RX J0058.2$-$7231 were found in the ASCA archive; 
both are found to have hard X-ray spectra (Yokogawa et al.\ 1999). 
Since the hard X-ray spectra are one of the signatures of XBPs, 
we extensively searched for pulsations, but found none, 
mainly due to their limited statistics. 
In fact, the total photon numbers for RX J0051.9$-$7311 and RX J0058.2$-$7231 
in the 1.5--6.0 keV band 
are respectively 361 and 273 (including background counts), 
which are only $\sim 1/3$ and $\sim 1/4$ of that of AX J0051$-$733.

RX J0103.2$-$7209 = 1SAX J0103.2$-$7209 is located in the radio source SNR 
0101$-$724 and has been found to exhibit coherent pulsations of 345.2 s 
with BeppoSAX  (Israel et al.\ 1998).  
This source was seen in the three ASCA archival data sets. 
Coherent pulsations of 348.9 s were found in the data set obtained 
on 1996 May 21, 
the highest statistics from this source among the three observations 
(Yokogawa, Koyama\ 1998). 

RX J0106.2$-$7205 showed no detectable X-ray emission in two ASCA
observations, while RX J0052.9$-$7158 has not been observed with ASCA.

In summary, the brightest two X-ray sources exhibit X-ray pulsations, 
while three weaker sources show no detectable pulsations.  
This latter fact is most probably due to their limited statistics. 
The remaining source was not observed with ASCA.  
Since Be-XBPs often exhibit X-ray outbursts,
we may have a chance to detect X-ray pulsations even from the three
weaker sources during their outburst episodes.
Thus, long-term monitoring over a wide area of the SMC would be very
fruitful in establishing the statistics of the Be-XBP population in a
neighboring galaxy. \par
\vspace{1pc}\par
We are very grateful to an anonymous referee for critical reading and 
useful comments.
We would like to thank all members of the ASCA team.
The Einstein and ROSAT data were obtained through the High Energy
Astrophysics Science Archive Research Center Online Service,
provided by the NASA/Goddard Space Flight Center. 
J.Y.\  is supported by JSPS Research Fellowship for Young Scientists.

\begin{figure}
%\begin{center}
\psfig{file=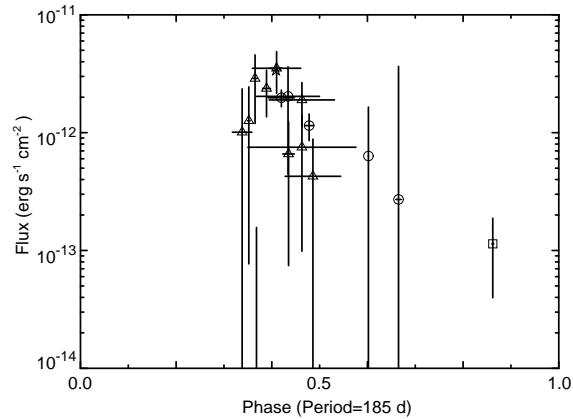,width=0.9\columnwidth}
%\end{center}
\caption
{Same as figure 4, but the data are folded with a trial period of 185 d. }
\end{figure}

\section*{References}
\re
Burke B.E., Mountain R.W., Harrison D.C., Bautz M.W., Doty J.P., Ricker G.R., Daniels P.J.\ 1991, IEEE Trans. ED-38, 1069
\re
Corbet R.H.D.\ 1984, A\&A 141, 91
\re
Cowley A.P., Schmidtke P.C., McGrath T.K., Ponder A.L., Fertig M.R., Hutchings J.B., Crampton D.\ 1997, PASP 109, 21
\re
Hughes J.P., Smith R.C.\ 1994, AJ 107, 1363
\re
Israel G.L., Stella L., Campana S., Covino S., Ricci D., Oosterbroek T. 1998, IAU Circ. 6999
\re
Kahabka P., Pietsch W.\ 1996, A\&A 312, 919
\re
Kahabka P., Pietsch W., Filipovi\'{c} M.D., Haberl F.\ 1999, A\&AS 136, 81
\re
Makishima K., Tashiro M., Ebisawa K., Ezawa H., Fukazawa Y., Gunji S., Hirayama M., Idesawa E. et al.\  1996, PASJ 48, 171
\re
Mathewson D.S., Ford V.L., Visvanathan N.\ 1988, ApJ 333, 617
\re
Nagase F.\ 1989, PASJ 41, 1
\re
Ohashi T., Ebisawa K., Fukazawa Y., Hiyoshi K., Horii M., Ikebe Y., Ikeda H., Inoue H. et al.\ 1996, PASJ 48, 157
\re
Schmidtke P.C., Cowley A.P., Crane J.D., Taylor V.A., McGrath T.K., 
Hutchings J.B., Crampton D.\ 1999, AJ 117, 927
\re
Serlemitsos P.J., Jalota L., Soong Y., Kunieda H., Tawara Y., Tsusaka Y., Suzuki H., Sakima Y. et al.\ 1995, PASJ 47, 105
\re
Tanaka Y., Inoue H., Holt S.S.\ 1994, PASJ 46, L37
\re
Wang Q., Wu X.\ 1992, ApJS 78, 391
\re
Yokogawa J., Koyama K.\ 1998, IAU Circ. 7009
\re
Yokogawa J., Imanishi K., Tsujimoto M., Nishiuchi M., Koyama K., Nagase F., 
Corbet R.H.D.\ 1999, ApJ submitted 

\begin{table*}[t]
\small
\begin{center}
Table~1.\hspace{4pt}Best-fit parameters to the spectra of AX J0051$-$733.\\
\end{center}
\vspace{6pt}
\begin{tabular*}{\textwidth}{@{\hspace{\tabcolsep}
\extracolsep{\fill}}p{6pc}ccccc}
\hline\hline\\ [-6pt]
Phase	& Photon index$^{\star}$	& Column density$^{\star}$		& Flux$^{\dagger}$			& Luminosity$^{\dagger\ddagger}$& $\chi^2$/$d.o.f.$ \\
	&		& (H~cm$^{-2}$)	& (erg~s$^{-1}$~cm$^{-2}$)		& (erg~s$^{-1}$)	& \\
\hline
Average		& 1.0$^{+0.4}_{-0.4}$	& 4.4$^{+7.1}_{-4.4}$ $\times$10$^{21}$	& 3.4$\times$10$^{-12}$		& 1.6$\times$10$^{36}$		& 36.04/33 \\
(0.0--1.0)	& 			& 			& 				& 				& 	\\
High		& 1.1$^{+0.3}_{-0.2}$	& 4.4 $\times$10$^{21}$	(fix)	& 4.3$\times$10$^{-12}$		& 2.0$\times$10$^{36}$		& 31.37/19 \\
(0.0--0.5)	&		& 			& 				& 				& 	\\
Low		& 0.8$^{+0.5}_{-0.4}$	& 4.4 $\times$10$^{21}$	(fix)	& 2.5$\times$10$^{-12}$		& 1.2$\times$10$^{36}$		& 12.52/12 \\
(0.5--1.0)	&		& 			& 				& 				& 	\\
\hline
\end{tabular*}
\vspace{6pt}
\par\noindent
$\star$: Errors are at 90 \% confidence level.
\par\noindent
$\dagger$:~In the 0.7--10.0 keV band.
\par\noindent
$\ddagger$:~Absorption-corrected value at the source distance of 60 kpc
 (Mathewson et al.\ 1988).
\end{table*}

\end{document}